\documentclass[12pt]{article}
\usepackage{graphicx}
\usepackage{amsfonts,amssymb,eucal,amsmath}
\begin{document}
 \parskip 0pt
 \parindent 24pt

\begin{center}
{\bf \Large Relativistic invariant projectors on a complex spinor
space and a rule of polarizations summation in a complex bispinor
space}
\\[0.5cm]
H. V. Grushevskaya\footnote{to be publishied in the Proceedings of
the 5th International Conference "Boyai-Gauss-Lobachevsky: Methods
of Non-Euclidean Geometry in Modern Physics", Minsk, Belarus,
October 10-13, 2006}
\\
{\it Physics Department, Belarusan State University,\\
    4 Nezavisimosti ave., Minsk 220030, BELARUS}
\\
    E-mail: grushevskaja@bsu.by
\end{center}

\begin{center}
 \bf Abstract
\end{center}
\begin{quotation}
Relativistic invariant projectors of states in a complex bispinor
space on a complex spinor space are constructed. An expression for
sections of bundle with connection on group \textsf{SU(4)} in an
explicit form has been obtained. Within the framework of the
proposed geometrical approach the rule of summation over
polarizations of states in a complex bispinor space has been
derived. It has been shown that states in a complex bispinor space
always describe a pair of Dirac's particles.
\end{quotation}

\section{Introduction}



Technique of covariant projection operators, which for the first time was
offered in paper \cite{fedorov}, is effectively used to calculate amplitudes of
scattering in quantum  field theories describing particles with a half-integer spin
\cite{Bogush}, \cite{Andreev} and vector-bosons \cite{Grush-vector-boson}.
In this method the projection operators are represented as matrixes - diads,
for such a construction it is necessary to determine  a set of basic bispinors.
As a rule, in quantum electrodynamics the basic sets describing states with
certain parity \cite{Gribov} are utilized.
In the paper we shall determine a basic set of states in a complex bispinor
space.

The goal of the paper is to construct relativistic invariant projectors of
states in a complex bispinor space on a complex spinor space and to study their
properties.

\section{
Projection operators on a spinor subspace}

Let us describe a particle by bispinor wave function,  components of
which are spinors $\xi,\ \dot{\xi}$ \cite{Umezawa}:
\begin{eqnarray}
  \Psi  \sim \left(
\begin{array}{c}
\xi\\
\dot \xi
\end{array}
\right)=\left(
\begin{array}{c}
\xi_1\\
\xi_2\\
\dot \xi_1\\
\dot \xi_2
\end{array}
\right)
 .\label{wave-bispinor-func1}
\end{eqnarray}
The spinors $\xi,\ \dot{\xi}$ are transformed by the representation of the
Lorentz group \cite{Gribov}:
\begin{eqnarray}
\xi = \left[\sqrt{
{p_0 +m\over 2m}}+(\vec \sigma \cdot\vec n)
\sqrt{ {p_0 -m\over 2m}}\right]\varphi e^{-\imath px},\label{wave-bispinor-func}
\\
\dot \xi = \left[\sqrt{
{p_0 +m\over 2m}}-(\vec \sigma \cdot\vec n)
\sqrt{ {p_0 -m\over 2m}}\right]\varphi e^{-\imath px} .\label{wave-spinor-func1}
\end{eqnarray}

Let us introduce spinor wave functions with  defined parity:
\begin{equation}
\begin{split}
 \Psi_1={1\over 2} (\xi+\dot \xi ),\quad
       \Psi_2={1\over 2} (\xi-\dot \xi ),
\end{split} \label{parity-wave-bispinor-func1}
\end{equation}
where $ \Psi_1$ is a spinor having positive parity,
$ \Psi_2$ is a spinor having negative parity.
Substituting eqs. (\ref{wave-bispinor-func}) and 
(\ref{wave-spinor-func1}) into 
eq.~(\ref{parity-wave-bispinor-func1}) we find these spinors in the
explicit form:
\begin{equation}
\begin{split}
 \Psi_1=\sqrt{{p_0 +m\over 2m}}  \varphi e^{-\imath px},\quad
       \Psi_2=(\vec \sigma \cdot\vec n)
\sqrt{ {p_0 -m\over 2m}} \varphi e^{-\imath px}.
\end{split}\label{parity-wave-spinor-func}
\end{equation}
Introducing into consideration symmetric and antisymmetric spinors we can
describe a Dirac particle restricting only appropriate spinor spaces being
subspaces of the bispinor space. It means that the bispinor space is divided into
two spinor subspaces: symmetric and antisymmetric spinor subspaces
${\cal H}^s$, ${\cal H}^a$:
\begin{eqnarray}
{\cal H}={\cal H}^s \oplus {\cal H}^a.
\end{eqnarray}
Therefore the bispinor wave function in ${\cal H}^s$ should be chosen in the
form:
\begin{eqnarray}
\Psi ^- ={1\over\sqrt{2m}}\left(
\begin{array}{c}
 \sqrt{ p_0 +m  }  \varphi_ {\lambda_ {s }}\\
     (\vec \sigma \cdot\vec n) \sqrt{  p_0 -m } \varphi _ {\lambda_{ a }}
\end{array}
\right)e^{-\imath px}= \nonumber \\
=\sqrt{ p_0 +m \over 2m }\left(
\begin{array}{c}
   \varphi_ {\lambda _{s }}\\
     (\vec \sigma \cdot \vec p/ (  p_0+m  )) \varphi _ {\lambda_{ a }}
\end{array}
\right)e^{-\imath px}=\left(
\begin{array}{c}
u^{\lambda _{s}}(p)\\
u^{\lambda _a}(p)
\end{array}
\right)    e^{-\imath px}.\nonumber \\
\label{parity-wave-spinor-func1}
\end{eqnarray}
Here the label $s(a)$ refers to a spinor with positive (negative) parity.
Normalization condition for a bispinor $u^{\lambda _s}$
immediately follows from the spinor form:
\begin{eqnarray}
& \overline u_\alpha ^{\lambda _s}(p) u_\alpha ^{\lambda _s'}(p)
   =\frac{1}{2}
   \left( \sqrt{p_0+m}\varphi^\dagger _{\lambda _s},
-\varphi^\dagger _{\lambda_s}
(\vec \sigma \cdot \vec n)\sqrt{p_0-m}
\right)
\left(\begin{array}{c}
\sqrt{p_0+m}\varphi _{\lambda _s'}\\
(\vec \sigma \cdot \vec n)\sqrt{p_0-m} \varphi  _{\lambda _s'}
\end{array}
\right)\nonumber \\[0.3cm]
& =(p_0+m- p_0+m)\delta _{\lambda_s \lambda _s'}/2m
            = \delta _{\lambda_s \lambda _s'}.
\label{Normalization}
\end{eqnarray}
As usual a summation is understood by twice meeting indexes.
Since the states with a given polarization can be considered as a basis
of two-dimensional spinor space, choosing in space-time tetrades
$s_\mu ^{\tau}$, $\tau =1,\ldots,4$  in the bispinor space the basis will
consist of
states which are determined by the equations:
\begin{eqnarray}
        \left({1\over 2}\gamma_5 \diagup \hspace{-3.5mm} s^{\tau} -\lambda  \right)u ^{\tau}=0.
\label{rel-basis0}
\end{eqnarray}
Let us choose as
$s_\mu ^{\tau}$ that have three zero components from four ones in the rest.
Then the basis of bispinor space in the rest has to be chosen by the following
form
\begin{eqnarray}
 u_\alpha \left(p,s^{1(2)} \right)=    {1\over \sqrt{2m}}\left(\begin{array}{c}
\sqrt{p_0+m}\varphi _{{1\over 2}(-{1\over 2})}\\
0
\end{array}
\right)\nonumber  \\[0.3cm]
 u_\alpha \left(p,s^{3(4)} \right)={1\over \sqrt{2m}}\left(\begin{array}{c}
                  0        \\
(\vec \sigma \cdot \vec n)\sqrt{p_0-m }\varphi _{{1\over 2}(-{1\over 2})}
\end{array}  \right) ,\ p_0=m     \label{rel-basis-vectors0}         .
\end{eqnarray}
Here one took into account that if a center-of-mass of a system is the rest
reference
system then the component $p_0$ of
four-dimensional velocity $p$ can be considered as the energy
$E=p_0$ of the particle in rest:
$E=p_0=m$.

One introduces
projection operators $P(s)$ with the help of relation

%
\begin{equation}
P(\lambda)\equiv P(\vec s)={1+ \vec \sigma \cdot \vec n\over 2}.
 \label{rest-spinor-projec-operator}
\end{equation}
The operator
$P(s)$ (\ref{rest-spinor-projec-operator})  in relativistic invariant form
is determined by the expression:
\begin{eqnarray}
P(s)\equiv {1+\gamma_5 \diagup \hspace{-3.5mm}  s\over 2}.
\label{spinor-projection-operator}
\end{eqnarray}
It follows from here that the helicity $\vec \sigma \cdot \vec n$ of a
particle along direction
$\vec s$ coincides with the direction
$\vec n$ up to  a sign:
\begin{equation}
\vec s=\pm \vec n \label{helicity}.
\end{equation}
A summation of projection operator over helicities gives unity
\begin{equation}
\sum_{\pm|\vec s|} P(\vec s)=1 \label{project-oper-sum},
\end{equation}
as it should be.

It is evidently that the spinor description is unsatisfactory because there
is no basic antisymmetric spinors in range of positive values  for
parameter
 $p_0$. Let us assume that the parameter
$p_0$ takes on the values: 
\begin{equation}
 |p_0|\ge m . \label{complexification}
\end{equation}
Let us make the change $p_0\to -p_0$ in the range of negative values  for
parameter
$p_0$.
Then one can find basic antisymmetric spinors
$u^+_\alpha \left(p,s^{\tau} \right)$, $\tau =1,\ldots,4$ \cite{Gribov}:
\begin{equation}
\begin{split}
 u^+_\alpha \left(p,s^{1(2)} \right)=    {\pm\imath\over \sqrt{2m}}
 \left(\begin{array}{c}
\sqrt{p_0-m}\varphi _{{1\over 2}(-{1\over 2})}\\
0
\end{array}
\right),  \\[0.3cm]
 u^+_\alpha \left(p,s^{3(4)} \right)={\pm\imath\over \sqrt{2m}}\left(\begin{array}{c}
                  0        \\
(\vec \sigma \cdot \vec n)\sqrt{p_0+m }\varphi _{{1\over 2}(-{1\over 2})}
\end{array}
 \right). 
\end{split}
  \label{antirel-basis-vectors0}
\end{equation}
It follows from here that the basic antisymmetric spinors are imaginary.
Hence, we have complexified the spinor space  by the change $p_0\to -p_0$.
Really, in a moving reference frame the change $p_0\to -p_0$ becomes
the change $p_0\to -p_0,\ \vec p\to -\vec p$. As a result of this change,
the symmetric spinor $\Psi ^- $ (\ref{parity-wave-spinor-func1}) becomes
an antisymmetric imaginary spinor $\Psi ^+ $:
\begin{eqnarray}
\Psi ^+ 
= {\pm \imath\over\sqrt{2m}}\left(
\begin{array}{c}
 \sqrt{ p_0 -m  }  \varphi_ {\lambda_ {s }}\\
     (\vec \sigma \cdot\vec n) \sqrt{  p_0 +m } \varphi _ {\lambda_{ a }}
\end{array}
\right)e^{\imath px}=\left(
\begin{array}{c}
u^{\lambda _{s}}(-p)\\
u^{\lambda _a}(-p)
\end{array}
\right)    e^{\imath px}.
\label{antiparity-wave-spinor-func1}
\end{eqnarray}

Now taking into account that only two nonzero basic spinors are
included into a basic set
(\ref{rel-basis-vectors0}), (\ref{antirel-basis-vectors0})
we  with the help of projection operator technique are able to write
 a rule of summation over spinor polarizations in the relativistic
  invariant form:
\begin{eqnarray}
\sum_{\lambda=\pm 1/2}  u_\alpha ^\lambda(p) \overline u_\beta^{\lambda }(p)
={1\over 2m }\sum_{\lambda=\lambda_s,\lambda_a}
(\diagup \hspace{-3.5mm}  p+m )_{\alpha \gamma}
u_\gamma ^\lambda \overline u_\beta^{\lambda }
=\sum_{\lambda=\pm 1/2} \sum_{\Phi}
\left( {\diagup \hspace{-3.5mm}  p+m\over 2m }\right)_{\alpha
\kappa}
P_{\kappa \delta} (\lambda)
|\Phi\rangle \gamma_0 \langle \Phi|
 P_{\delta \beta}(\lambda ) \nonumber \\[0.2cm]
=\sum_{\tau=1}^2 \sum_{\Phi'} \left( {\diagup \hspace{-3.5mm}  p+m\over 2m }
\right)_{\alpha \kappa}
P_{\kappa \delta} (s^\tau)
|\Phi'\rangle \gamma_0 \langle \Phi'|
 P_{\delta \beta}(s^\tau )\nonumber\\
   \label{rel-basic-sum001}
\end{eqnarray}

Now, one will take into account a unity decomposition over vectors
$|\Phi' \rangle $: 
\begin{eqnarray}
\hat 1=\sum_{\Phi'}  |\Phi' \rangle \gamma_0 \langle \Phi'|
\label{rel-basic-set0},
\end{eqnarray}
where the matrix - diad $|\Phi' \rangle \gamma_0 \langle \Phi'|\equiv
\left.\left(u^+_\alpha \cdot \gamma_0 ({u^+_\alpha }^T)^\dag \right)
\right|_{p_0=-m}$, $u^+_\alpha $ is defined by the formula
(\ref{antirel-basis-vectors0}), $\varphi ^T_{-1/2}=(0\ 1)$ ,
$\varphi ^T_{+1/2}=(1\ 0)$ , $\dag$ denotes a complex conjugation,
$\hat 1$ is the unity operator.
Then eq.~(\ref{rel-basic-sum001}) can be rewritten as 
\begin{eqnarray}
\sum_{\lambda=\pm 1/2}  u_\alpha ^\lambda(p) \overline u_\beta^{\lambda }(p)
={1\over 2m }\sum_{\lambda=\lambda_s,\lambda_a}
(\diagup \hspace{-3.5mm} p+m )_{\alpha \gamma}
u_\gamma ^\lambda \overline u_\beta^{\lambda }
=\sum_{\tau= 1}^2
\left( {\diagup \hspace{-3.5mm}  p+m\over 2m }\right)_{\alpha \kappa}
P_{\kappa \beta} (s^\tau) \hat 1  = {1\over 2m }
   (\diagup \hspace{-3.5mm}  p+m )_{\alpha \beta}.
 \label{rel-basis-sum0}
\end{eqnarray}

In a similar manner as for the symmetric spinors we find a sum of
antisymmetric spinor products over polarizations:
\begin{eqnarray}
 \sum_{\lambda_s=
\pm 1/2}  u_\alpha ^{\lambda_s}(-p) \overline u_\beta^{\lambda _s}(-p)=
{1\over 2m }\sum_{\lambda=\lambda_s,\lambda_a}  (\diagup \hspace{-3.5mm}  p-m )_{\alpha \kappa}
u_\kappa ^\lambda \overline u_\beta^{\lambda } =\nonumber \\
=\sum_{\tau=3}^4 \sum_{\Phi'} \left( {\diagup \hspace{-3.5mm}  p-m\over 2m }\right)_{\alpha \kappa}
P_{\kappa \delta} (s^\tau)(\pm\imath )
|\Phi'\rangle \gamma_0 \langle \Phi'|(\pm \imath )
 P_{\delta \beta}(s^\tau )  \nonumber \\[0.2cm]
=\sum_{\tau= 3}^4
\left( {\diagup \hspace{-3.5mm}  p-m\over 2m }\right)_{\alpha \kappa}
P_{\kappa \beta} (s^\tau)
(-\hat 1)=
{1\over 2m }
   ( m- \diagup \hspace{-3.5mm}  p )_{\alpha \beta}.
 \label{antipolarization-sum001}
\end{eqnarray}

\section{
Projection operators on bispinor space }
Let us examine a bispinor space generated by the bispinors
$\Psi$ defined by the expression
(\ref{wave-bispinor-func1}), upper components of which are the spinors
 $ \xi $ defined by the expression (\ref{wave-bispinor-func}), and 
lower components are $\dot \xi $ defined by the expression
(\ref{wave-spinor-func1}).

Let $s_\mu ^{\tau}$, $\tau =1,\ldots ,4$ in
eq.~(\ref{rel-basis0}) has 
three zero components from four ones.
Then, in rest reference system one has to choose as a basis of
bispinor space the following bispinors
\begin{eqnarray}
 u_\alpha \left(p_0,\vec p=0,s^{1(2)} \right)=
{1\over \sqrt{2m}} \left(\begin{array}{c}
\sqrt{p_0+m}\varphi _{{1\over 2}(-{1\over 2})}\\
0
\end{array}
\right)   \label{rel-basis-vectors1}, \\[0.3cm]
 u_\alpha \left(p_0,\vec p=0,s^{3(4)} \right)=
{1\over \sqrt{2m}}\left(\begin{array}{c}
                  0        \\
\sqrt{p_0+m }\varphi _{{1\over 2}(-{1\over 2})}
\end{array}  \right)      \label{rel-basis-vectors2};
\end{eqnarray}
where bispinors $u_\alpha \left(p,s^{1(2)} \right)$ are transformed over
the representation $\xi$, and 
bispinors $u_\alpha \left(p,s^{3(4)} \right)$ are transformed over
the representation
$\dot{\xi} $.
The tetrad $s$ is called the helicity also.
From the expressions (\ref{wave-bispinor-func}),  
(\ref{wave-spinor-func1}), and (\ref{helicity}) 
describing  $u_\alpha \left(p,s^{\tau} \right)$
in an arbitrary reference frame 
one has the equality:
\begin{eqnarray}
\left(\vec \sigma \cdot \vec s^{1(2)} u_\alpha\right)(p_0,\vec p=0,s^{1(2)} )=
\pm
\kappa u_\alpha(p_0,\vec p=0,s^{1(2)} ), \label{projection1} \\
\left(\vec \sigma \cdot \vec s^{3(4)} u_\alpha\right)(p_0,\vec p=0,s^{3(4)} )=
\pm
(-\kappa )u_\alpha(p_0,\vec p=0,s^{3(4)} ), \label{projection2}
\end{eqnarray}
where $\kappa$ is determined by the expression 
\begin{equation}
\kappa =\sqrt{{p_0-m\over p_0+m}} \label{eigenvalue-spin}
\end{equation}
One concludes from here that
$\kappa $ must be equal to 
$\pm 1$: $\kappa =\pm 1$.
But $\kappa (p_0=m)\neq 1$. 
It means that although  the bispinors with helicity $s^\tau ,$
$\tau=1,2$ under the condition 
$p_0=m$, are solutions of the Dirac's equation
but they are not solutions of equations for eigenvalues of relativistic spin
of  Dirac particle.
Hence, in the region
$p_0\geq m$ for values  of  the parameter 
$p_0$ there do not exist 
 basis of the bispinor space.
To find the basis and, hence, to construct the bispinor space one
extends the region for values  of  the parameter
$p_0$:
\begin{equation}
 |p_0|\ge 0   \label{complexification1}
\end{equation}

Let us consider the region $|p_0|\leq m$. 
According to the expression (\ref{eigenvalue-spin}) one has to make the change
$\vec n \to \imath \vec n$
in this region. From here it follows that one can define  
a complex bispinor $\breve{u}$ as 
\begin{eqnarray}
 \breve{u} = \left(
\begin{array}{c}
\left(\sqrt{
{p_0 +m\over 2m}}+\imath (\vec \sigma \cdot\vec n)
\sqrt{ {p_0 -m\over 2m}}\right) \varphi_{\lambda_+}\\
\left( \sqrt{
{p_0 +m\over 2m}}-\imath (\vec \sigma \cdot\vec n)
\sqrt{ {p_0 -m\over 2m}}\right) \varphi_{\lambda_-}
\end{array}
\right)
=\left(\begin{array}{c}
 \breve{u}^{\lambda_+}\\
 \breve{u}^{\lambda_-}
\end{array} \right)
. \label{system-bispinor}
\end{eqnarray}
Let us find its  Dirac conjugation $\overline{ \breve{u}}$.
To do it 
one utilizes a probabilistic interpretation of scalar product of bispinor
wave functions
 $ \breve{u} $, $\overline{ \breve{u}}$. 
It means that we have to compose a conserved quantity.
Evidently, a product of spinors
$\xi,\ \dot{\xi}$ is a relativistic invariant because they are transformed on
different representations of  Lorentz group.
Really, we have
\begin{eqnarray}
     \dot  \xi ^\prime {^\dagger}   {\xi '}  ={\dot{\xi}} ^\dagger
e^{-{\chi \over 2}(\vec \sigma \cdot \vec n)}
e^{{\chi \over 2}(\vec \sigma \cdot \vec n)}\xi= {\dot{\xi}}^\dagger \xi .
\end{eqnarray}
Writing  $\xi,\ \dot{\xi}$ through linear combinations of symmetric and
antisymmetric spinors
$\Psi_1,\ \Psi_2$, one gets that 
\begin{eqnarray}
{\dot{\xi}}^\dagger \xi =(\Psi_1^\dagger - \Psi_2^\dagger ,\Psi_1+ \Psi_2)
= \Psi_1^\dagger \Psi_1 - \Psi_2^\dagger \Psi_2 =
\overline u^\lambda(p)u^\lambda(p). \label{conjugated-breve-bispinor}
\end{eqnarray}
It follows from here that
$\overline u^\lambda(p)u^\lambda(p)$ is the relativistic invariant.
In the probabilistic interpretation the product
$\overline{\breve{u}}^\lambda(p)\breve{u}^\lambda(p)$ 
plays the same role in  bispinor space as the product
$\overline u^\lambda(p)u^\lambda(p)$ in spinor space.
Hence, the product
$\overline{ \breve{u}}^\lambda(p)\breve{u}^\lambda(p)$ must be
relativistic invariant.
To satisfy this condition, in accordance with the expression
(\ref{conjugated-breve-bispinor}) the probability 
$\overline{ \breve{u}}^\lambda(p)\breve{u}^\lambda(p)$
must be represented as a combination of terms ${\dot{\xi}}^\dagger \xi $.
It means that the  conjugated complex bispinor $\overline{ \breve{u}}$
can be defined as
\begin{eqnarray}
\overline{ \breve{u} }=\left[\gamma_5
\left(
\begin{array}{c}
\varphi
^\dagger\left(\sqrt{
{p_0 +m\over 2m}}-\imath(\vec \sigma \cdot\vec n)
\sqrt{ {p_0 -m\over 2m}}\right)  \\
\varphi
^\dagger\left( \sqrt{
{p_0 +m\over 2m}}+\imath (\vec \sigma \cdot\vec n)
\sqrt{ {p_0 -m\over 2m}}\right)
\end{array}
\right)\right]^T .
\end{eqnarray}
Let us calculate a norm of bispinor
$\breve u$ introduced with the help of the relation
(\ref{system-bispinor}):
\begin{eqnarray}
\overline {\breve{u}}_\alpha {\breve{u}}_\alpha=
\left(
\begin{array}{c}
\varphi_{\lambda_+}^\dagger\left(\sqrt{
{p_0 +m\over 2m}}+\imath(\vec \sigma \cdot\vec n)
\sqrt{ {p_0 -m\over 2m}}\right)  \\
\varphi_{\lambda_-}^\dagger\left( \sqrt{
{p_0 +m\over 2m}}-\imath (\vec \sigma \cdot\vec n)
\sqrt{ {p_0 -m\over 2m}}\right)
\end{array}
\right)^T 
\left(
\begin{array}{c}
\left(\sqrt{
{p_0 +m\over 2m}}+\imath(\vec \sigma \cdot\vec n)
\sqrt{ {p_0 -m\over 2m}}\right) \varphi_{\lambda_+}\\
\left( \sqrt{
{p_0 +m\over 2m}}-\imath(\vec \sigma \cdot\vec n)
\sqrt{ {p_0 -m\over 2m}}\right) \varphi_{\lambda_-}
\end{array}
\right)=2 .
\end{eqnarray}

Now let us show that in the built space of complex bispinors there exist
a basis satisfying the  system from the Dirac equation and
the equation for relativistic spin which describes a Dirac particle.
Hermitian conjugated Dirac equation has the form:
\begin{eqnarray}
\breve u^{\dagger}(\diagup \hspace{-3.5mm}   p^{\dagger}-m)=0
\label{moment-Dirak-ermit} ,
\end{eqnarray}
where 
\begin{eqnarray}
 \diagup \hspace{-3.5mm}   p^{\dagger}=
 (\gamma_0p_0-\vec \gamma\cdot \vec p)^{\dagger} =-(
 \gamma_0p_0+\vec \gamma\cdot \vec p)
\label{ermit-moment0}
\end{eqnarray}
owing 
to hermicity of $\gamma_0 $ and 
anti-hermicity of $\gamma_i $.
Let us multiply eq.~(\ref{moment-Dirak-ermit}) by 
$\gamma_5 $ from right. 
Then
taking into account that
the matrix $\gamma_5$ is represented in the form:
$\gamma_5 =
\imath \gamma _0 \gamma _ 1 \gamma _2 \gamma _3$,
one gets 
\begin{eqnarray}
 \breve u^\dag (\gamma_0 p_0(\imath \gamma_0\gamma_1 \gamma_2 \gamma_3 )
+ \vec \gamma \cdot \vec p(\imath \gamma_0 \gamma_1 \gamma_2 \gamma_3 )+m)=0.
 \label{breve-Dirak}
\end{eqnarray}
The matrix $\gamma_5$ is a pseudoscalar:
$\gamma_5=-{\imath \over 4!} \epsilon _{\mu \nu \sigma \rho}
\gamma ^\mu \gamma ^\nu \gamma ^\sigma \gamma ^\rho =
\imath \gamma _0 \gamma _ 1 \gamma _2 \gamma _3$.
Using anticommutation relations for the $\gamma$ - matrixes :
\begin{eqnarray}
 \gamma_\mu \gamma_\nu  + \gamma_\nu \gamma_\mu =2\delta_{\mu\nu},
\label{ermit-moment}
\end{eqnarray}
one can rewrite eq.(\ref{breve-Dirak}) as
\begin{eqnarray}
 \breve u^\dag \left(-{\imath \over 4!} \epsilon _{ \nu \sigma \rho \mu}
 \gamma ^\nu \gamma ^\sigma \gamma ^\rho \gamma ^\mu \gamma_0 p_0
-{\imath \over 4!} \epsilon _{\nu \mu  \sigma \rho}
\gamma ^\mu \gamma ^\nu \gamma ^\sigma \gamma ^\rho \gamma_1 p_1
+{\imath \over 4!} \epsilon _{\sigma \mu \nu    \rho}
\gamma ^\mu \gamma ^\nu \gamma ^\sigma \gamma ^\rho \gamma_2 p_2 \right.
\nonumber \\
\left.
-{\imath \over 4!} \epsilon _{\rho \mu \nu   \sigma }
\gamma ^\mu \gamma ^\nu \gamma ^\sigma \gamma ^\rho \gamma_3 p_3
+m\right)=0.
 \label{breve-Dirak1}
\end{eqnarray}
Therefore the new field $\overline{ \breve{u} }$ satisfies the Dirac
equation
\begin{eqnarray}
\overline {\breve u}(\diagup \hspace{-3.5mm}   p + m)=0
\label{Dirak-ermit1}.
\end{eqnarray}

Let us put $p_0=0$.
Then the basic bispinors
$u^\tau_\alpha (0)$, $\tau =1,\ldots, 4$ are defined in the following way:
\begin{equation}
\begin{split}
{\breve{u}}^1(0)={1\over \sqrt{2}}\left(
\begin{array}{c}
1\\
0\\
0\\
0
\end{array}
\right),\
{\breve{u}}^2(0)={1\over \sqrt{2}}\left(
\begin{array}{c}
0\\
1\\
0\\
0
\end{array}
\right),\
{\breve{u}}^3(0)={1\over \sqrt{2}}\left(
\begin{array}{c}
0\\
0\\
1\\
0
\end{array}
\right),
\
{\breve{u}}^4(0)={1\over \sqrt{2}}\left(
\begin{array}{c}
0\\
0\\
0\\
1
\end{array}
\right).
\end{split}
\label{basis_bispinors}
\end{equation}
We see that the center-of-mass system
($p_0=m,\ \vec p=0$) 
is excluded as the rest frame.
Changing $s$ for 
$\pm \imath n$ with the help of the expression 
(\ref{helicity}), the system of equations 
(\ref{projection1}, \ref{projection2}) can be rewritten 
in the form
\begin{eqnarray}
(\vec \sigma \cdot \vec n {\breve{u}}^{1(2)}_\alpha )(0)=
{\breve{u}}^{1(2)}_\alpha(0), \label{projection1-1} \\
(\vec \sigma \cdot \vec n {\breve{u}}^{3(4)}_\alpha ) (0)=
- {\breve{u}}^{3(4)}_\alpha (0 ) \label{projection2-1}.
\end{eqnarray}
Hence, we prove that the projection of particle spin in built complex bispinor
space is described by the operator of Dirac particle spin projection.
Using the technique of calculations with the help of projection operators
determined in our case by the expressions
(\ref{spinor-projection-operator}) and taking into account that
all 4 basic bispinors are nonzero we get a rule of summation over
polarization in the form 
\begin{eqnarray}
 \sum_{\lambda_+=\pm 1/2}  {\breve{u}}_\alpha ^{\lambda_+}(p)
\overline {\breve{u}}_\beta^{\lambda_+ }(p)=
\sum_{\vec s=\{\pm \vec n \}} \sum_{\Phi} P_{\alpha \delta }(\vec s)
|\Phi \rangle \gamma_5\langle \Phi |
P_{\delta \beta}(\vec s)
=
\sum_{\tau=1}^4 \sum_{\Phi '}P_{\alpha \delta }( s^\tau)
|\Phi ' \rangle \gamma_5
\langle \Phi '|P_{\delta \beta } . \label{begin-of-polarization-sum0}
\end{eqnarray}
According to the expression (\ref{basis_bispinors}) we have  
\begin{eqnarray}
\sum_{\Phi '} |\Phi ' \rangle \gamma_5 \langle \Phi '| ={\hat 1\over 2}
\label{operator-unity}
\end{eqnarray}
because the matrix - diad $|\Phi' \rangle \gamma_5 \langle \Phi'| $ is equal to
\begin{eqnarray}
|\Phi' \rangle \gamma_5 \langle \Phi'|\equiv
\left.\left\{\left[\imath\ \Im m\ {\breve u}^+_\alpha\right] \cdot \gamma_5
\left[\imath\ \Im m\ \left(({\breve u}^+
_\alpha )^T\right)^{\dag}\right]
\right\}
\right|_{p_0=0},
\end{eqnarray}
 $\breve u^+_\alpha $ is defined by the formula
(\ref{system-bispinor}).

We see that due to the expression
(\ref{basis_bispinors}), the sum over polarizations 
(\ref{begin-of-polarization-sum0}) is built on 
a basic half-set:
 $\sum_{\Phi} |\Phi \rangle \gamma_5\langle \Phi |={\hat {1}\over 2}$. 
Besides, the norm of complex spinor is two times less than the norm of
complex bispinor.
Hence, the probability to find a quantum object described by the complex
bispinor is two times larger than for a quantum object described by the complex
spinor.
From here it follows physical sense of the basis with
$p_0=0$. 
The built basis describes a pair "particle-antiparticle".\
As we saw for the particle $p_0=E$, for the antiparticle 
$p_0=-E$.  
Therefore the parameter $p_0 $ of the pair "particle-antiparticle"\
equals to zero.

Now,  for the basic bispinors
${\breve{u}}^{1(2)}_\alpha (p)$, we can rewrite the sum
(\ref{begin-of-polarization-sum0}) of bispinors 
${\breve{u}}_\alpha ^{\lambda_+}(p)$ transformed
by representation  $\xi $ without point 
in the relativistic invariant form
\begin{eqnarray}
 \sum_{\lambda_+ =\pm 1/2}  {\breve{u}}_\alpha ^{\lambda_+}(p)
 \overline {\breve{u}}_\beta^{\lambda _+}(p)=
{1\over 2m }\sum_{\tau=1}^4  (\diagup \hspace{-3.5mm}   p+m )_{\alpha \gamma}
{\breve{u}}_\gamma ^\tau \overline {\breve{u}}_\beta^{\tau } /2 \nonumber \\
=
{1\over 2m }\sum_{\tau=1}^4  (\diagup \hspace{-3.5mm} p+m )_{\alpha \gamma}
P_{\gamma \beta}(s^\tau )/2
= {1\over 2}\left({\diagup \hspace{-3.5mm}  p+m \over 2m }\sum_{\tau=1}^4
 P(s^\tau ) \right)_{\alpha \beta}.
 \label{polarization-sum01}
\end{eqnarray}
Since $P(s)$ is a projection operator one obtains
\begin{eqnarray}
\sum_{\tau =1}^4  P(s^{\tau  })/2= 1.
\label{rel-basis-sum}
\end{eqnarray}
Substituting the expression
(\ref{rel-basis-sum}) into the expression 
(\ref{polarization-sum01}), we get the rule of summation
over polarizations $\lambda_+$:
\begin{eqnarray}
 \sum_{\lambda_+=\pm 1/2}  {\breve{u}}_\alpha ^{\lambda_+}(p)
 \overline {\breve{u}}_\beta^{\lambda _+}(p)=
   \left({\diagup \hspace{-3.5mm}  p+m \over 2m}\right)_{\alpha \beta} .
 \label{polarization-sum1}
\end{eqnarray}
 Similarly for the basic bispinors
 ${\breve{u}}^{3(4)}_\alpha (p)$,
one gets the sum over polarizations for bispinors transformed by
representation $\dot \xi$:
\begin{eqnarray}
 \sum_{\lambda_-=\pm 1/2}  {\breve{u}}_\alpha ^{\lambda_-}(p)
\overline {\breve{u}}_\beta^{\lambda_- }(p)=
\sum_{\vec s=\{\pm \vec n \} } \sum_{\Phi} P_{\alpha \delta }(\vec s)
(\imath)|\Phi \rangle \gamma_5\langle \Phi |(\imath)
P_{\delta \beta}(\vec s) = \nonumber \\
=- \sum_{\tau=1}^4 \sum_{\Phi '}P_{\alpha \delta }( s^\tau)
|\Phi ' \rangle \gamma_5
\langle \Phi '| P_{\delta \beta }
=-{\hat 1\over 2}
\sum_{\tau=1}^4 {\breve{u}}_\alpha  ^\tau (p)
\overline {\breve{u}}_\beta^{\tau } (p) 
=\nonumber \\
=-{1\over 2m }\sum_{\tau=1}^4  (\diagup \hspace{-3.5mm} p-m )_{\alpha \gamma}
{\breve{u}}_\gamma ^\tau \overline {\breve{u}}_\beta^{\tau }/2 
=-
{1\over 2m }\sum_{\tau=1}^4  (\diagup \hspace{-3.5mm} p-m )_{\alpha \gamma}
P_{\gamma \beta}(s^\tau )/2
=  \left(
{ m- \diagup \hspace{-3.5mm}  p \over 2m } \right)_{\alpha \beta}.
 \label{polarization-sum02}
\end{eqnarray}

The projection operators
${\breve{u}}_\alpha ^{\lambda }(p)
\overline {\breve{u}}_\beta^{\lambda }(p)$ 
constitute a total set.
Indeed, it is easy to show that 
\begin{eqnarray}
 \sum_{\lambda=\{\lambda_\pm\}}  {\breve{u}}_\alpha ^{\lambda}(p)
\overline {\breve{u}    }_\beta^{\lambda} (p)
=  \left(
{ \diagup \hspace{-3.5mm} p +m \over 2m } \right)_{\alpha \beta}+\left(
{ m- \diagup \hspace{-3.5mm} p \over 2m } \right)_{\alpha \beta}=\hat 1.
 \label{projection-operatos-sum}
\end{eqnarray}
Let us find the projection operators
 $\pi_{\alpha \beta }(p)\equiv {\breve{u}}_\alpha (p)
\overline {\breve{u}}_\beta (p)$
in a explicit form. To do it, we rewrite 
the expression (\ref{polarization-sum02}) in the form: 
\begin{eqnarray}
 \sum_{\lambda_-=\pm 1/2}  {\breve{u}}_\alpha ^{\lambda_-}(p)
\overline {\breve{u}    }_\beta^{\lambda_-} (p)
=- { 1 \over 4m }\sum_{\vec s=\{\pm \vec n \}} \left[
( \diagup \hspace{-3.5mm} p -m )\left(1- \gamma_5 \vec \gamma \cdot \vec s
\right)\right]_{\alpha \beta}.
 \label{explicit-projection-operator}
\end{eqnarray}
Then, we obtain
\begin{eqnarray}
  \pi_{\alpha \beta }^\lambda (p)\equiv{\breve{u}}_\alpha ^\lambda (p)
\overline {\breve{u}}_\beta ^\lambda (p)
= -{ 1 \over 4m } \left[
(\diagup \hspace{-3.5mm} p -m)\left(1-\gamma_5 \vec \gamma \cdot \vec s \right)\right]_{\alpha \beta}.
 \label{explicit-projection-operator1}
\end{eqnarray}
Replacing
$p\to -p$, 
interchanging upper and lower components of bispinors by multiplying on left
and on right by matrix
$\gamma_5$, and using rules of multiplication for matrixes
$\gamma_5, \gamma_\mu $, the expression
(\ref{polarization-sum01}) becomes an expression of the form:
\begin{eqnarray}
 \sum_{\lambda_- =\pm 1/2}  {\breve{u}}_\alpha ^{-\lambda_-}(-p)
 \overline {\breve{u}}_\beta^{-\lambda_- }(-p)
={1\over 4m}\left[
(\diagup \hspace{-3.5mm} p +m  )\sum_{\vec s=\{\pm \vec n \}}
\gamma_5 \left(1- \gamma_5 \vec \gamma \cdot \vec s \right)
 \gamma_5 \right]_{\alpha \beta}.
 \label{explicit-antiprojection-operator}
\end{eqnarray}
It follows from here that there exist yet one type of projection operators
which is denoted as
 $\pi_{\alpha \beta }^{-\lambda}(-p)$:
\begin{eqnarray}
 \pi_{\alpha \beta }^{-\lambda}(-p)\equiv {\breve{u}}_\alpha ^{-\lambda}(-p)
\overline {\breve{u}}_\beta ^{-\lambda}(-p)=
 {1\over 4m}\left[
(\diagup \hspace{-3.5mm} p +m  )  \left(1-   (\vec \gamma \cdot \vec s) \gamma_5 \right)
 \right]_{\alpha \beta}.
 \label{explicit-antiprojection-operator1}
\end{eqnarray}

Having in mind the explicit expression, for example, of the projection
operator
 ${\breve{u}}_\alpha ^{\lambda }(p)\overline {\breve{u}}_\beta^{\lambda }(p)$,
we can establish  correspondence between the complex spinor and complex
bispinor representations by the following way.
From the expression (\ref{explicit-projection-operator1}) for
the projection operator 
and from the fact that complex bispinor describes the fermion pair,
it follows that given projector projects on a discrete space with a basis
from eigenvectors, eigenvalues of which lay in a gap of energy spectrum of
the Dirac's operators. Therefore
integer spin $\vec s$  can be attributed to these vectors.
Hence, physical states are described by such bispinors
${\breve{u}}_\alpha ^{\lambda }(p)$,
the projection of which equals to zero:
\begin{eqnarray}
(  \pi^\lambda _{\alpha \beta } {\breve{u}}_\beta ) (p)
= 0 .
\end{eqnarray}
Therefore owing the motion equations
the physical states of quantum system 
must be  solutions of the equation as
\begin{eqnarray}
(  \pi^\lambda _{\alpha \beta } {\breve{u}}_\beta ) (p)
= -{ 1 \over 4m }  \left[\left(
( \diagup \hspace{-3.5mm} p -m )\breve{u}^\lambda (p)\right)_\alpha
-\left(
( \diagup \hspace{-3.5mm}  p -m ) \gamma_5 \vec \gamma \cdot \vec s \right)_{\alpha \beta}
{\breve{u}}^\lambda_\beta (p)  \right] \nonumber\\
={ 1 \over 4m } \left(
( \diagup \hspace{-3.5mm} p -m ) \gamma_5 \vec \gamma \cdot \vec s \right)_{\alpha \beta}
{\breve{u}}^\lambda_\beta (p)=0 \nonumber \\
 \label{bispinor-motion-equation}
\end{eqnarray}
or solutions of the equation as
\begin{eqnarray}
(  \pi^{-\lambda} _{\alpha \beta } {\breve{u}}_\beta ) (-p) 
= { 1 \over 4m }  \left[\left( ( \diagup \hspace{-3.5mm} p  +m )\breve{u}^{-\lambda} (-p)\right)_\alpha
-\left(
( \diagup \hspace{-3.5mm} p   +m )  (\vec \gamma \cdot \vec s) \gamma_5\right)_{\alpha \beta}
{\breve{u}}^{-\lambda}_\beta (-p)  \right]=\nonumber\\
= - { 1 \over 4m } \left(
( \diagup \hspace{-3.5mm} p   +m )  (\vec \gamma \cdot \vec s) \gamma_5\right)_{\alpha \beta}
{\breve{u}}^{-\lambda}_\beta (-p)=0. \nonumber \\
 \label{bispinor-motion-equation1}
\end{eqnarray}
From here we get that there  exist the following relation  between the
spinors
 $u^\lambda_\alpha (p)$, $v^\lambda_\alpha (p)\equiv u^{-\lambda}_\alpha (-p)$
and 
the bispinors
$\breve{u}^\lambda_\alpha (p)$, $\breve{ u}^{-\lambda}_\alpha (-p)$,
respectively: 
\begin{eqnarray}
u^\lambda (p) =
\gamma_5 (\vec \gamma \cdot \vec s)
{\breve{u}}^\lambda  (p),\\
v^\lambda  (p) =
 (\vec \gamma \cdot \vec s)\gamma_5
{\breve{u}}^{-\lambda}  (-p).
\end{eqnarray}
But, 
the complex description in spinor space is equivalent
to a real description in Minkowski space \cite{Penrous}.
Now, we can show that gauge-invariant states of system lay in  a connection
$\varpi $ of bundle
associated with principal bundle, a base of which is the Minkowski space
$\mathbb{M}$, and the connection is the unitary group
\textsf{SU (4)}. 
Indeed, we see from eqs.~(\ref{bispinor-motion-equation}) and 
 (\ref{bispinor-motion-equation1}) 
that the projection $\pi_{\alpha \beta}$ of fiber for this bundle
on the base
$\mathbb{M}$ equal to zero
gives a motion equation of Dirac's particle.

Further, we consider  sections of found bundle.

\section{Section of bundle with connection on group
\textsf{SU(4)}}
Let us examine a complex spinor
$|x_\lambda ^{R(L)} \rangle $ 
with a given helicity.
Here  indexes $R,L$ denote 
right- and left-helical Dirac's particles, respectively.
Since the complex bispinor describes a doubled spinor, one can write in rest
the identity:
\begin{eqnarray}
\sum_\mu |\xi_\mu \left>\gamma_5 \right< \xi_\mu|\psi \rangle
\equiv \sum_{\lambda} \left( |x_\lambda ^{L}
\rangle \gamma_0 \langle x_\lambda ^{L} |
+ |x_\lambda ^{R} \rangle \gamma_0\langle
x_\lambda ^{R} |\right)\psi \rangle .
\label{proection-0}
\end{eqnarray}


Let us show that the identity
(\ref{proection-0}) in a rest reference frame 
can be written as
\begin{eqnarray}
\sum_\mu |\xi_\mu \left> \gamma_5 \right< \xi_\mu|\psi \rangle + {1\over
2}\sum_\lambda \left[ |\xi_\kappa \left> \gamma_5\left(\sigma^+_\lambda
\right)_{\kappa \nu} \right< \xi_\nu|\psi \rangle + |\xi_\kappa
\left> \gamma_5\left(\sigma^-_{\lambda} \right)_{\kappa \nu} \right<
\xi_\nu|\psi \rangle \right]
\nonumber \\
\equiv \sum_{\lambda}\left[ {1\over 2}\left( |x_\lambda ^{L}
\rangle  \gamma_0\langle x_\lambda ^{L} |
+ |x_\lambda ^{R} \rangle  \gamma_0 \langle
x_\lambda ^{R} |\right)\psi \rangle +{1\over 2}\left( |x_\lambda
^{L} \rangle  \gamma_0 \langle x_\lambda ^{L} |+ |x_\lambda ^{R} \rangle
 \gamma_0 \langle x_\lambda ^{R} |\right)\psi \rangle
\right],
\label{proection}
\end{eqnarray}
where matrices
$\sigma^\pm_{ \lambda}$ 
are determined by
\begin{eqnarray}
\sigma^+_\lambda =\epsilon _{\lambda ij} \sigma_{ij};\
\sigma^-_{\lambda} =\epsilon _{\lambda ji} \sigma_{ij}, \ \lambda
, i,j =1,2,3; \label{4dPaulimatrix}
\end{eqnarray}
indexes
$\kappa, \mu , \nu $ 
run over
$s, \dot{s} $; $\sigma_{\mu \nu}= {\imath \over
2}[\gamma_\mu , \gamma_\nu ]$, $\gamma_\mu$ 
are Dirac matrices
.
Hereinafter,  we will omit sign $ \pm $. Since the skew-symmetric
tensor $\epsilon _ {\lambda jk} $ appears in the equation for $\sigma _ {\lambda} $,
the generalized Pauli matrixes  $ \sigma _{\lambda} $ are  pseudo-vectors.


It follows from the expansion of the wave function  $ | \psi
\rangle $ (\ref {proection}) in a series   that   the projection $P$  defined by
the expression:
\begin{eqnarray}
P|\psi\rangle = {1\over 2}\left( |x_\lambda ^{R} \rangle \gamma_0\langle
x_\lambda ^{R} |+ |x_\lambda ^{L} \rangle \gamma_0 \langle x_\lambda ^{L}
|\right)\psi \rangle ,
 \label{proection1}
\end{eqnarray}
can be represented in the form
\begin{eqnarray}
P|\psi\rangle ={1\over 2}\left( |x_\lambda ^{R} \rangle \gamma_0\langle
x_\lambda ^{R} |+ |x_\lambda ^{L} \rangle \gamma_0 \langle x_\lambda ^{L}
|\right)\psi \rangle
={1\over 2} \left\{ \sum_\mu |\xi_\mu \left> \gamma_5\right< \xi_\mu|\psi
\rangle + \sum_\lambda |\xi_\kappa \left> \gamma_5\left(\sigma_\lambda
\right)_{\kappa \nu} \right< \xi_\nu|\psi \rangle \right\}
\label{proection2}.
\end{eqnarray}

Let us prove  that the projector (\ref {proection2}) selects
states with a given orientation. The right side of the expression
(\ref{proection2}) in the representation $ \{| \xi_i \rangle \}_ i$ is written as
\begin{eqnarray}
P\langle \xi _\lambda |\psi \rangle ={1\over 2} \left\{ 1 +
\sum_\lambda \langle \xi _\lambda |\xi_\kappa \left>\gamma_5
\left(\sigma_\lambda \right)_{\kappa \nu}
\right.
\right\} \langle \xi _\nu |\psi \rangle \label{proection3}
\end{eqnarray}
%
We can introduce a four-vector $s^\mu =\{0,s^\lambda \},\
s ^\lambda = \langle \xi _ \lambda
| \xi_\kappa \rangle $ describing a spin of the system. A
convolution of the vector $s ^\lambda $  with matrixes $
\sigma_\lambda $ is situated on the right side in Eq.~(\ref
{proection3}). It follows from here, that in $s $-representation
$P=P (s) $ is determined by the expression
\begin{eqnarray}
P(s)={1\over 2}(1+ \gamma_5 s^\mu \gamma_\mu )    \qquad  s^\mu =\{0,s^\lambda \}.
\label{proector(s)}
\end{eqnarray}
The formula (\ref{proector(s)}) is the known expression
for the projection operator selecting spinors with a given orientation in
rest.
%
After left multiplication  on $ \langle x_\lambda ^R | $ the requirement
of orthonormal basis $
\langle x_{\lambda_i} ^R | x_{\lambda_j} ^R \rangle =\delta_{ij} $
allows us to rewrite the expression
(\ref{proection2}) in rest in the form
\begin{eqnarray}
  \langle x_\lambda ^{R}
|x_\lambda ^{L} \rangle \gamma_0\langle x_\lambda ^{L} |\psi \rangle
= \langle x_\lambda ^{R}|\xi_\kappa \left> \left(\gamma_5  \sigma_\lambda
\right)_{\kappa \nu} \right< \xi_\nu |x_\lambda ^{L} \rangle \gamma_0
\langle x_\lambda ^{L}|\psi \rangle \label{proection5-0}
\end{eqnarray}
Let us introduce the designations
\begin{eqnarray}
x_\lambda = \langle x_\lambda ^{R}| x_\lambda ^{L} \rangle;\qquad
\xi_\kappa = \langle x_\lambda ^{R}|\xi_\kappa \rangle .
\label{reduction}
\end{eqnarray}
Taking into account  the expression
(\ref{reduction}),  after simple transformations
the expression
(\ref{proection5-0})
is written as
\begin{eqnarray}
   x_\lambda
 = \xi_\kappa ^\dag \left(\sigma_\lambda \gamma_5 \right)_{\kappa \nu}
\xi_\nu  \label{proection5}
\end{eqnarray}

Now we can calculate
the square $\left| x \right|^2 =x^\lambda x_\lambda $  of module of the
expression (\ref {proection5}). Since
\begin{eqnarray}
x^\lambda
 = \xi_\kappa^\dag \left(\gamma_5 ({\sigma_\lambda^T})^\dag \right)_{\kappa \nu}
\xi_\nu . \label{proection5-1}
\end{eqnarray}
we have:
\begin{eqnarray}
 \left| x \right|^2
  =
 \left(\
\begin{array}{c}
 \xi_s\\
\xi_{\dot{s
}}
\end{array}
 \right)^{\dagger}
\sum_{i=1}^3 \gamma_5 ({\gamma_i^T})^\dag  \gamma_i
 \left[
\xi^\kappa (\xi^\dag \gamma_5)_\kappa
 \right]
\left(  \begin{array}{c}
 \xi_s\\
\xi_{\dot{s
}}
\end{array}
 \right)
 = |\xi |^2\left(\
\begin{array}{c}
 \xi_s\\
\xi_{\dot{s
}}
\end{array}
 \right)^{\dagger}
 \left(
\begin{array}{cc}
0 &  1\\
1& 0
\end{array}
 \right)
\left(  \begin{array}{c}
 \xi_s\\
\xi_{\dot{s
}}
\end{array}
 \right)=|\xi |^4
. \label{square-proection5}
\end{eqnarray}
Here $ | \xi | ^2 =\sum _ {\nu =1} ^4 \xi^2_\nu $,
Dirac representation of matrices $ \gamma ^ \mu, \ \mu =0,1,2,3 $  is
chosen.
It follows from here that the bispinors representation is two-valued one.

\section{Conclusion}

Relativistic invariant projectors of states in a complex bispinor
space on a complex spinor space are constructed. The last allows to
find an expression for sections of bundle with connection on group
\textsf {SU(4)} in an explicit form. Within the framework of the
proposed geometrical approach the rule of summation over
polarizations of states in a complex bispinor space has been
derived. It has been shown that states in a complex bispinor space
always describe a pair of Dirac's particles.

\end{document}